\documentclass [prl,twocolumn,tightenlines,amssymb,showpacs]{revtex4}
\usepackage{amsmath}
\usepackage {graphicx}
\begin{document}
\title{Temperature Evolution of Sodium Nitrite Structure in a
Restricted Geometry}
\author{A.V. Fokin, Yu.A. Kumzerov, N.M. Okuneva}\affiliation{Ioffe Phys.-Tech. Institute, 26
Politekhnicheskaya, 194021, St.-Petersburg, Russia}
\author{A.A. Naberezhnov}\email {alex.nabereznov@pop.ioffe.rssi.ru} \affiliation{Ioffe Phys.-Tech.
Institute, 26 Politekhnicheskaya, 194021, St.-Petersburg, Russia}
\author{S.B. Vakhrushev}\email {s.vakhrushev@pop.ioffe.rssi.ru} \affiliation{Ioffe Phys.-Tech.
Institute, 26 Politekhnicheskaya, 194021, St.-Petersburg, Russia}
\author{I.V. Golosovsky, A.I. Kurbakov}\affiliation{Petersburg Nuclear Physics Institute, 188300,
Gatchina, Leningrad distr., Russia}
\date{\today}
\begin{abstract}
The NaNO$_{2}$ nanocomposite ferroelectric material in porous
glass was studied by neutron diffraction. For the first time the
details of the crystal structure including positions and
anisotropic thermal parameters were determined for the solid
material, embedded in a porous matrix, in ferro- and paraelectric
phases. It is demonstrated that in the ferroelectric phase the
structure is consistent with bulk data but above transition
temperature the giant growth of amplitudes of thermal vibrations
is observed, resulting in the formation of a "premelted state".
Such a conclusion is in a good agreement with the results of
dielectric measurements published earlier.
\end{abstract}

\maketitle
\bigskip
\section{INTRODUCTION}
\label{intro} The problem of the physical properties of materials
in a restricted geometry is one of the "hot" points of modern
solid state physics and is not only of fundamental interest but
also of practical importance. Along with films, filaments etc.
there is large and important group of restricted geometry objects,
namely materials confined within porous media (hereinafter we will
call them confined materials - CM). In recent years properties of
CM and in particular various types of phase transitions (PT)
(superconducting \cite{p1,p2}, superfluid \cite{p3,p4},
melting-freezing \cite{p5,p6,p7,p8,p9,p10,p11} and others PTs
\cite{p12,p13,p14,p15,p16,p17,p18,p19,p20,p21,p22} in different CM
have been extensively studied by different experimental methods
including calorimetry \cite{p5,p7,p20}, NMR \cite{p9,p21},
ultrasonic \cite{p8,p9} and dielectric \cite{p12,p14,p15}
measurements, Raman \cite{p10,p13}, X-ray \cite{p12,p16,p17,p18}
and neutron scattering \cite{p10,p14,p22,p23,p24,p25,p26},
differential thermal analysis \cite{p19} etc. It has been shown
that CM can form either a system of isolated particles \cite{p10}
or a net of interconnected dendrite clusters \cite{p14} and their
physical properties differ drastically from those in corresponding
bulk samples and strongly depend on different characteristics of
porous matrices and embedded substances such as pore size and
geometry, wetting ability, surface tension, interaction between CM
and surface of host matrix and so on.

Finite-size effects in ferroelectrics were observed for the first
time at the beginning of 1950s \cite{p16}. It was shown that the
physical properties of dispersed ferroelectrics are significantly
different from those of the bulk materials in particular when the
characteristic size becomes comparable with correlation length of
order parameter critical fluctuations. In detail the modern
theoretical and experimental situation is well described in the
review \cite{p27} and references therein. During last years the
development of new nanotechnologies gave the strong impetus to the
study of ferroelectric microcomposites as a new basis of
ferroelectric memories or an active component in fine-composite
materials, however the principal attention was devoted to the thin
films or granular ferroelectrics. On the other hand the very
interesting and surprising results were recently obtained for
ferroelectric CM. In particular the dielectric measurements of
NaNO$_{2}$, KH$_{2}$PO$_{4}$ (KDP) and Rochelle salt confined in
different porous matrices have shown \cite{p14,p15,p28} the growth
of dielectric permittivity $\varepsilon $ above temperature of
ferroelectric phase transition T$_{c}$ for all materials and all
matrices and unexpected increase of T$_{c}$ at decreasing of
characteristic size D for KDP. The most remarkable result was the
giant growth of $\varepsilon $ (up to 10$^{8}$ at 100~Hz - the
record value!) at approaching to the bulk melting temperature
(T$_{melt} $= 557~K) that was observed for NaNO$_{2}$ embedded
into an artificial opal matrix \cite{p15}. The temperature
dependence of $\varepsilon $ in CM case essentially differs from
analogous dependence for the bulk NaNO$_{2}$ \cite{p29} typical
for classical ferroelectrics, and this giant growth of dielectric
permittivity was attributed to the extremely broadened melting
process \cite{p15}, but no experimental evidence was presented.

We have attempted to study the temperature evolution of structure
of confined NaNO$_{2}$ in porous glass with pores size 7 nm at
temperatures below and above T$_{c}$ by the method of neutron
diffraction to clarify the microscopic origin of observed
anomalies of dielectric permittivity. This method was successfully
used for study of structure evolution of water, cyclohexane
\cite{p23,p24,p25,p26} and liquid mercury \cite{p10} confined
within porous media at melting-freezing PT, but no diffraction
study of microcomposite ferroelectric materials (except our
preliminary results \cite{p14}) were performed earlier. Moreover
we do not know any paper dealing with the detailed structure
refinement (including determination of thermal parameters) of any
confined solid materials.

\section{RESULTS}
\label{sec:1} 
\begin{figure}
\includegraphics[width=\columnwidth,clip=] {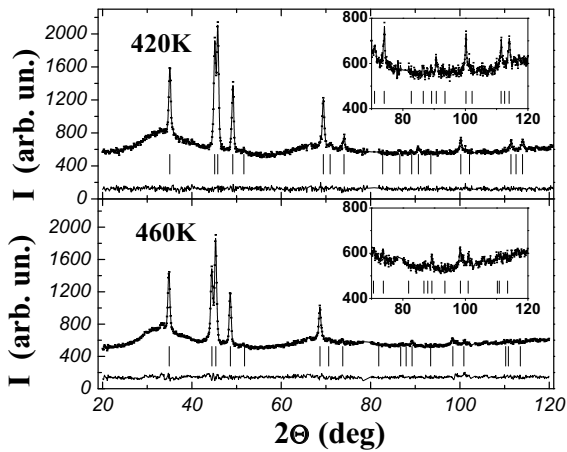}
\caption{Neutron diffraction patterns for NaNO$_{2}$ in porous
glass at 420~K (ferroelectric phase) and 460~K (paraelectric
phase). The inserts present diffraction patterns at large Q in
enlarged scale.} \label{1}
\end{figure}

The samples were prepared by immersing of the
preliminary warmed up platelets of the porous glass in the melted
NaNO$_{2}$ in the sealed quartz container. The glass samples were
tested by the mercury intrusion porosimetry and the pore sizes
were found to be 7$\pm $1~nm. The volume amount of the salt was
about 25\%. Measurements were performed on the powder
diffractometer G4-2 (LLB, Saclay, France) at 2.3434~{\AA} at room
temperature (RT), 400~K, 420~K, 440~K, 450~K and 460~K, i.e. below
and above ferroelectric PT temperature T$_{c}$ $\approx$ 438~K.
The diffraction patterns for 420~K (ferroelectric phase) and 460~K
(paraelectric phase) are presented in Fig.~\ref{1}. The diffuse
background observed in addition to the normal diffraction peaks
due to porous silica glass was used to determine the nearest Si-O
and O-O distances. These parameters were found to be almost equal
to those for glass silicate tetrahedron SiO$_{4}$ \cite{p30} and
were practically temperature independent therefore the cavity
sizes in the host glass matrix do not depend on temperature.

The observed diffraction peaks corresponding to the orthorhombic
structure were slightly asymmetric with the width larger then the
instrumental resolution, but much smaller then the value expected
for the scattering on the isolated 7~nm particles. It leads to a
conclusion that due to high wetting ability the sodium nitrite
forms a kind of interconnected clusters probably of the dendrite
type. Their average size was found from structure refinement to be
about 45~nm and was practically temperature independent. One
should mention that the situation is quite different from the case
of non-wetting compounds like the liquid mercury, which forms on
cooling a system of independent particles with characteristic size
equal to the average pore diameter \cite{p10}.

The diffraction patterns in ferroelectric phase were fitted in the
frames of the \textbf{Im2m} space group and following the paper
\cite{p31} a pseudo-mirror plane perpendicular to \textbf{b} axis
was included at \textbf{y}=0 to take into account incomplete
ordering of the NO$_{2}$ groups. In this case the long-range order
parameter can be determined as \textbf{ç=f}$_{1}$\textbf{-f}$_{2}$
\cite{p31}, where the fraction \textbf{f}$_{1}$ of the NO$_{2}$
groups was placed on one side of the plane and
\textbf{f}$_{2}$\textbf{=1-f}$_{1}$ on the other side.

Below T$_{c}$ our results are in a good agreement with published
data for the bulk NaNO$_{2}$, however the anisotropic thermal
parameters $\beta_{ij}$ are slightly higher then for the bulk
\cite{p31}. The diffraction patterns above T$_{c}$ correspond to
the paraelectric phase with space group \textbf{Immm. } The
heating through T$_{c}$ results in the decrease of intensity of
most of the peaks. The observed effect is much stronger than that
in the case of bulk NaNO$_{2}$ \cite{p31} and is in agreement with
our earlier data \cite{p14}.

\section{DISCUSSION}
\label{sec:2}

The results of the refinement procedure have revealed two main
distinguishing features of the temperature evolution of structure
of embedded sodium nitrite.

\begin{figure}
\includegraphics[width=\columnwidth,clip=] {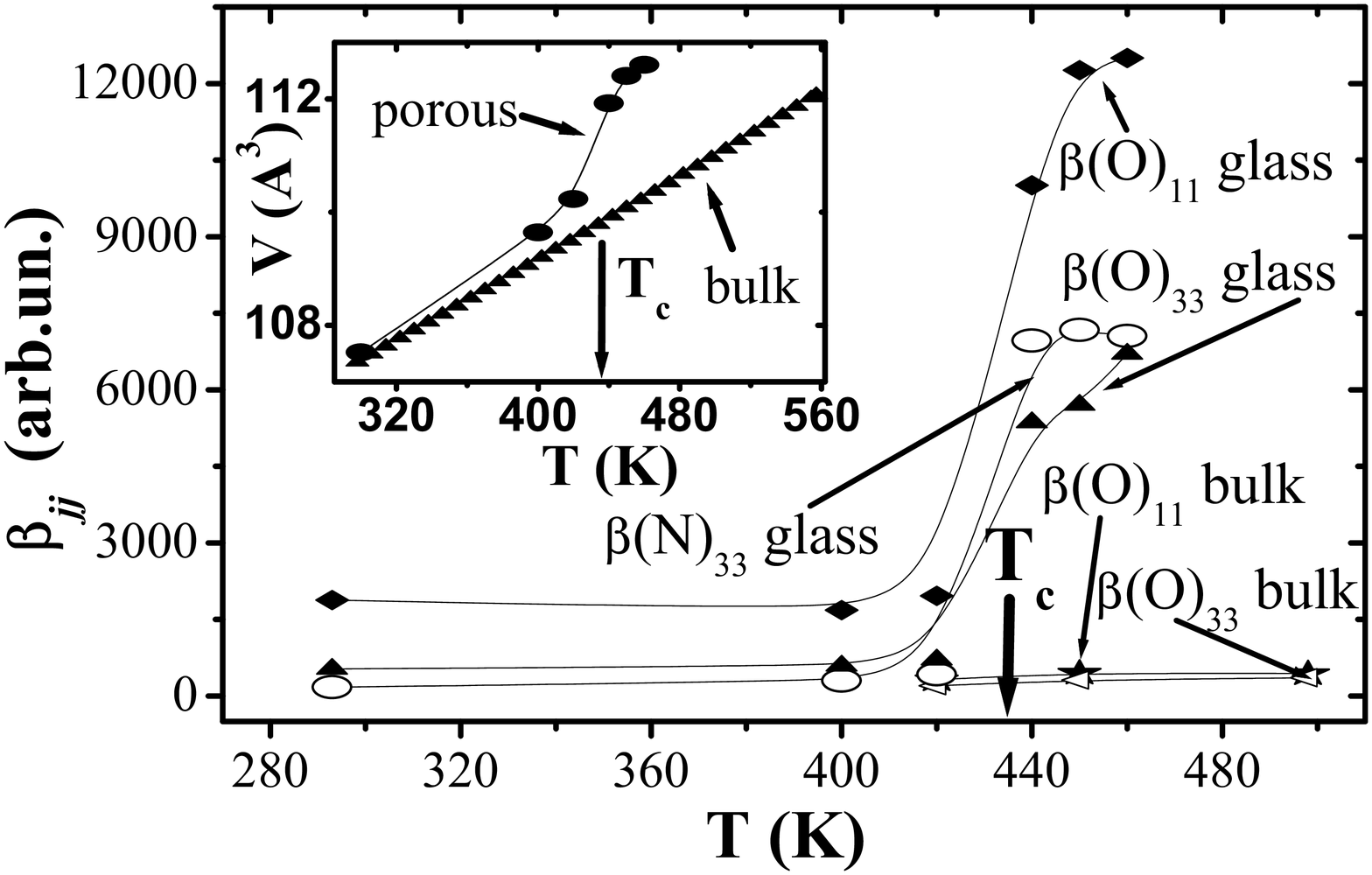}
\caption{Temperature dependences of thermal parameters $\beta
_{jj}$ for bulk \cite{p31} and porous samples. On insert
temperature dependences of unit cell volume for bulk \cite{p33}
and CM are presented. The errors do not exceed the size of
symbols. Vertical arrows indicate temperature of ferroelectric PT
for the bulk.} \label{2}
\end{figure}

The first one is the essential increase of elementary cell volume
upon passing through T$_{c}$ (insert on Fig.~\ref{2}). Here one
can note that a similar phenomenon was observed for the
overwhelming majority of different materials at melting
\cite{p32}. The detailed analysis of temperature dependences of
lattice parameters \textbf{a}, \textbf{b} and \textbf{c} shows
that in ferroelectric phase the confined NaNO$_{2}$ expands in the
\textbf{a} and \textbf{b} directions and contracts in the
\textbf{c} direction similar to the bulk material, but above
T$_{c}$ the lattice parameters \textbf{a} and \textbf{b} increase
rapidly and \textbf{c} decreases slowly then in the bulk. As far
back as 1961 S. Nomura \cite{p29} had shown that the misalignment
of the NO$_{2}$ anion is responsible for the anomalous thermal
expansion in the \textbf{b} direction (along macroscopic
polarization). He had also supposed that the expansion in the
\textbf{a} direction and the contraction in the \textbf{c}
direction could be explained by rotation or rotational vibration
about arbitrary axis of non-spherical NO$_{2}$ group, which holds
its long axis parallel to the \textbf{c} direction in the
equilibrium. Later K. Takahashi and W. Kinase \cite{p34} have
proposed the microscopic mechanism of the ferroelectric PT and
have shown that the mixed type of NO$_{2}$ group rotation around
the \textbf{a} and \textbf{c} axes is achieved. They have shown
also that in ferroelectric phase there are the potential barriers,
which have strong angular dependences and limit the rotation of
the NO$_{2}$ around \textbf{a} and \textbf{c} axes.

In terms of such a model the observed behavior of unit cell
parameters could be explained by the increase of these rotations
above T$_{c}$, experimentally displaying as a growth of thermal
vibrations of ions. And indeed, the second distinguishing feature
is the steep growth of the thermal parameters $\beta_{ij}$ (Fig.
\ref{2}) above T$_{c}$ pointing out the "looseness" (or softening)
of the structure.

\begin{figure}
\includegraphics[width=.8\linewidth,height=.7\linewidth, clip=]{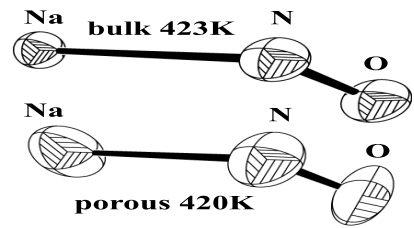}
\caption{50 per cent ellipsoids of thermal vibrations for bulk and
porous samples at 420~K (below T$_{c}$).} \label{3}
\end{figure}

\begin{figure}
\centering
\includegraphics[width=.8\linewidth,height=.7\linewidth, clip=]{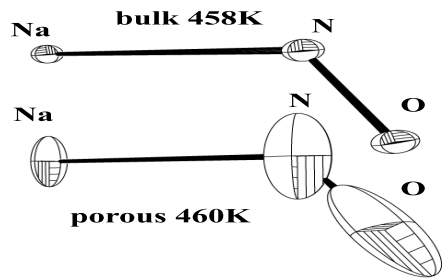}
\caption{5 per cent ellipsoids of thermal vibrations for bulk and
porous samples at 460~K (above T$_{c}$). \textbf{\underline
{Note}}: the ellipsoids of 5\% probability are presented, inasmuch
as oxygen thermal vibrations are very large for porous sample}
\label{4}
\end{figure}

 Using obtained $\beta_{ij}$, the thermal vibration ellipsoids
for constituent ions were constructed at all measured temperatures
and have been compared with those for the bulk material
\cite{p31}. The results of refinement at 420~K (below T$_{c}$) and
460~K (above T$_{c}$) are presented as ellipsoids of 50\%
probability in Fig. \ref{3} and as ellipsoids  of 5\% probability
in Fig. \ref{4} (inasmuch as oxygen thermal displacements are very
large for porous sample). For the bulk material these ellipsoids
are close to a sphere at all temperatures and their characteristic
sizes increase insignificantly on heating. For sodium nitrite in
porous glass below T$_{c}$ these ellipsoids are clearly
anisotropic and slightly larger then for the bulk, but on heating
through T$_{c}$ the shape and size of the thermal vibration
ellipsoids change drastically. In the paraelectric phase (above
T$_{c}$) the vibrations of Na and N form practically flat disks
perpendicular the \textbf{b} direction for Na and the \textbf{a}
direction for N as a result of mixed rotation around \textbf{a}
and \textbf{c} axes, while oxygen ions form very stretched
ellipsoids predominantly along the \textbf{a} and \textbf{c}
directions, as it should be expected at increasing of rotation
around \textbf{b} axis. The obtained values of oxygen thermal
displacements along the \textbf{c} and \textbf{a} directions at
460~K (above T$_{c}$) are equal to 1.21~{\AA} and 0.93~{\AA}
respectively (i.e. more than 25\% of O-O (3.34~{\AA}) distance for
neighboring NO$_{2}$ groups) and essentially exceed the Lindemann
criterion, which states that bulk material will melt when the
average value of thermal displacements of nuclei exceeds 10\% -
15\% of internuclear distances \cite{p32,p35,p36}.

Having the picture of ion thermal vibrations and keeping in mind
the results of structure refinement, we can suppose that the
"looseness" (or softening) of the structure is a distinctive (and
true intrinsic) feature of this CM corresponding to the formation
above T$_{c}$ (more then 100 degrees below the bulk T$_{melt}$)
"premelted state" initially manifesting itself in the oxygen
sublattice. In this case the mentioned above growth of dielectric
permittivity of NaNO$_{2}$ above T$_{c}$ \cite{p15} could be
explained by appearance of ionic current due to jumping diffusion
of constituent ions, firstly oxygen ones.

Early analogous premelted effects were studied in metasilicates
(Na$_{2}$SiO$_{3}$, Li$_{2}$SiO$_{3}$ \cite{p37,p38}) and some
others minerals \cite{p39}, such as diopside, anorthite etc.,
where the heat capacity anomalies have been observed beginning
from 100 to 200 degrees below relevant melting temperatures, and
for those it was demonstrated, that premelting represents
unquenchable configurational changes within phases remaining
crystalline up to congruent melting points \cite{p39}. It is
different from premelting of ice in porous glass \cite{p40} or in
exfoliated graphite \cite{p41} matrices, where the premelted state
is formed at the surface layer between the ice and host matrix and
strongly depends on surface curvature. In our case the "premelted
state" has a volume character and probably originates from some
size effect of yet unclear nature.

In conclusion, for the first time the details of the structure of
a solid material embedded in a porous matrix was determined by
neutron diffraction. The temperature evolution of structure in a
restricted geometry was studied for the ferroelectric NaNO$_{2}$
embedded in a porous glass and it was shown that this CM forms a
kind of interconnected clusters probably of the dendrite type with
practically temperature independent average size about 45~nm.
Above T$_{c}$ the volume "premelted" state is formed, manifesting
itself in a sharp growth of the thermal motion parameters,
softening of lattice and increasing of lattice volume. In such a
case the possible appearance of ionic current due to oxygen
jumping diffusion is proposed as a reason of observed giant growth
of dielectric permittivity. On cooling below T$_{c}$ macroscopic
polarization and potential barriers suppress the lattice softening
and the normal ferroelectric phase exists, but even at room
temperature the thermal vibrations are different from those for
the bulk material.

\bigskip

The work was supported by the RFBR (grants 00-02-16883,
01-02-17739), the Russian Program "Neutron Researches of Solids"
(the contract 01.40.01.07.04) and "Solid State Nanostructures"
(grant 99-1112).

\end{document}